\documentclass[]{spie}
\usepackage[]{graphicx}
\graphicspath{{figures/}}
\usepackage{color}

\title{Robo-AO Kitt Peak: Status of the system and deployment of a sub-electron readnoise IR camera to detect low-mass companions} 

\author{Ma\"issa Salama\supit{a}, Christoph Baranec\supit{a}, Rebecca Jensen-Clem\supit{b}, Reed Riddle\supit{b}, Dmitry Duev\supit{b}, Shrinivas Kulkarni\supit{b}, Nicholas M. Law\supit{c}
\skiplinehalf
\supit{a}Institute for Astronomy, University of Hawai'i at M\=anoa, Honolulu, HI 96822, USA; \\
\supit{b}Department of Astrophysics, California Institute of Technology, Pasadena, CA 91125, USA; \\
\supit{c}Department of Physics and Astronomy, University of North Carolina at Chapel Hill, Chapel Hill, NC 27599, USA
}

\authorinfo{Further author information: (Send correspondence to M.S.)\\M.S.: E-mail: msalama@ifa.hawaii.edu, Telephone: (808) 956-8007}

\begin{document} 
\maketitle

%%%%%%%%%%%%%%%%%%%%%%%%%%%%%%%%%%%%%%%%%%%%%%%%%%%%%%%%%%%%% 
\begin{abstract}
%Aim: What are you going to do ?
%Justification: Why are you doing it?
%Method: How are you going to do it?
%Results: What did you conclude?
We have started an initial three-year deployment of Robo-AO at the 2.1-m telescope at Kitt Peak, Arizona as of November 2015. We report here on the project status and two new developments with the Robo-AO KP system: the commissioning of a sub-electron readnoise SAPHIRA near-infrared camera, which will allow us to widen the scope of possible targets to low-mass stellar and substellar objects; and, performance analysis and tuning of the adaptive optics system, which will improve the sensitivity to these objects. Commissioning of the near-infrared camera and optimizing the AO performance occur in parallel with ongoing visible-light science programs.
\end{abstract}


%%%%%%%%%%%%%%%%%%%%%%%%%%%%%%%%%%%%%%%%%%%%%%%%%%%%%%%%%%%%%
\section{INTRODUCTION}
\label{sec:intro} 

Robo-AO \cite{Baranec14}$^,$ \cite{JOVE13} is a robotic laser guide star, adaptive optics (AO) instrument now exclusively deployed on the 2.1-m telescope at the Kitt Peak National Observatory (KPNO) in Arizona. The system is based on a reconfiguration of the Robo-AO prototype that operated on the 1.5-m telescope at the Palomar Observatory in California, from 2011 to 2015. Robo-AO has the unique ability to observe over 200 targets in one night at the visible diffraction limit, $\approx 0.1''-0.15''$. Robo-AO currently observes in the visible but a SAPHIRA \cite{Atkinson14}$^,$ \cite{Baranec15}$^,$ \cite{Atkinson16} detector infrared camera will be added as a visiting instrument to the system in order to widen the scope of possible targets as well as simultaneously image in the visible and the infrared.

The Robo-AO instrument is introduced in \S \ref{sec:instru} and its commissioning to Kitt Peak is described in \S \ref{sec:toKP}. We describe the tools used to determine observing conditions and image quality in \S \ref{sec:Tools}. In \S \ref{sec:Opt} we describe the optimizations and calibrations performed since Robo-AO has been deployed to Kitt Peak. The performance of the system is analyzed in \S \ref{sec:Perf} and the description and status of the infrared camera is explained in \S \ref{sec:IRcam}.

%%%%%%%%%%%%%%%%%%%%%%%%%%%%%%%%%%%%%%%%%%%%%%%%%%%%%%%%%%%%%
\section{Robo-AO Instrument} 
\label{sec:instru}

\begin{figure}[h!]
\begin{center}
\includegraphics[height=3in]{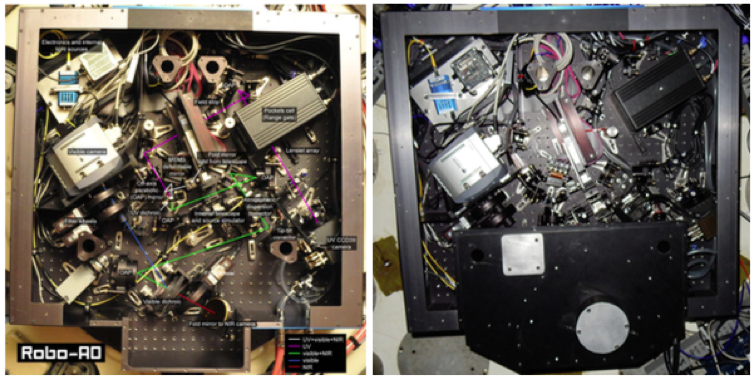}
\caption[fig:instrument]
{\label{fig:instrument} Images of the Robo-AO Cassegrain instrument at Palomar (\textit{left}) and at Kitt Peak (\textit{right}). The light from the telescope enters the instrument with a fold mirror and onto an OAP which images the pupil onto the MEMS deformable mirror (white line path in the Palomar image on the left). Then a UV dichroic splits the UV light (magenta), which passes through a condenser lens, onto an OAP, into the Pockels cell, through a lenslet array, and onto the Shack-Hartmann wavefront sensor while the visible and infrared light (green) passes through a second OAP relay, which includes the atmospheric dispersion corrector (ADC) and tip-tilt corrector (currently not in use, see text in \S \ref{sec:instru}). A visible dichroic then sends the visible light (blue) to the visible CCD camera and infrared light (red) to the soon-to-be deployed infrared camera. The large black platform at the bottom of the Kitt Peak image is the mount for the infrared camera.}
\end{center}
\end{figure}

 Robo-AO uses a Rayleigh scattering laser guide star with a line of sight focus at $\approx$10km. It uses a 10kHz pulsed 10W ultraviolet ($\lambda$ = 355nm) laser, and is invisible to the eye and thus safe for pilots. Figure \ref{fig:instrument} shows the Robo-AO Cassegrain instrument at Palomar and Kitt Peak. The light from the telescope enters the instrument with a fold mirror which reflects the light to an off-axis parabolic (OAP) mirror, which then images the pupil onto a MEMS (Micro-Electro-Mechanical) deformable mirror. After reflection, a UV dichroic splits the ultraviolet light from the laser which then passes through a condenser lens, to a field stop. An OAP then collimates the light, which passes through a Pockels cell controlling the range gate before the pupil is imaged onto an $11 \times 11$ lenslet array. A refractive relay then images the Shack-Hartmann pattern onto a UV optimized E2V CCD39 detector. The visible and infrared light that passes through the UV dichroic is refocused and sent to a second OAP relay which includes an atmospheric dispersion corrector (ADC) and tip-tilt corrector before being split by a visible dichroic to the visible and infrared cameras. Robo-AO is currently equipped with a $1024^2$ pixel EMCCD visible camera which formed a $44'' \times 44''$ field of view with a plate scale of $0.043''$/pixel at Palomar and now forms a $36'' \times 36''$ field of view with a plate scale of $0.035''$/pixel at Kitt Peak. Although there is a tip-tilt corrector, we are currently not using tip-tilt correction and instead are processing the images with post-facto shift and add techniques on data taken at 8.6Hz. 
 
%\vskip -0.25in

%%%%%%%%%%%%%%%%%%%%%%%%%%%%%%%%%%%%%%%%%%%%%%%%%%%%%%%%%%%%%
\section{Kitt Peak Commissioning}
\label{sec:toKP}
%\textcolor{magenta}{[Help: Christoph, Reed, Becky, Dima]}

Robo-AO was decommissioned from Palomar in June 2015, then in November 2015, less than six months later, was commissioned on the 2.1-m telescope in Kitt Peak, Arizona after having the optics reconfigured in the lab in October 2015. 

\subsection{Optics Reconfiguration}
\label{sec:realign}
In October 2015, we made several hardware changes to the Robo-AO instrument in preparation for the move to Kitt Peak, Arizona. In order to speed up the deployment to Kitt Peak and avoid reconfiguring the optics in the natural guide star (NGS) arm of the instrument, the f-number, F/8.6, was purposefully kept the same by installing a circular aperture mask on the primary mirror, reducing the diameter from 2.1-m to 1.85-m. Due to the different focal length of the Kitt Peak telescope, the $\approx 10km$ laser beacon comes to a focus at a different location, but since we are using the same pupil size and lenslet array, the optics in the wavefront sensor (WFS) needed to be realigned and the condenser lens was changed. For the telescope simulator, a new motorized stage was added to recenter the the source when doing calibrations of non-common path aberrations. A laser light source, which has a very narrow bandwidth like the on-sky laser but has resolved speckles which require a fiber scrambler, was used in addition to the LED light previously used at Palomar, which has a 30nm bandwidth and thus forms chromatic errors with the condenser lens and requires a narrowband filter which made it faint.

\subsection{Installation at Kitt Peak}
\label{sec:KPinstall}
The instrument was installed on the 2.1m telescope at Kitt Peak in November 2015 and several changes were made compared to its installation at Palomar. A new mounting adaptor was developed, and the instrument was rotated by 23.9$^o$ so that North points up and East points right on the EMCCD detector. Periscope mirrors were installed, which allow the laser beam to be moved closer to the telescope aperture and ensure that it does not get clipped by the dome opening. We initially placed the beam right next to the telescope aperture but we found that very low altitude Rayleigh backscatter was getting through the Pockels cell shutter while it was off. We then removed the periscope, but plan on re-installing it about 30cm from the telescope aperture in order to prevent excessive Rayleigh background light onto the WFS. The electronics packaging was modified into several smaller racks to balance the load on the telescope. The laser mount was changed to a three-point mounting plate to make alignment simpler and more robust, and several minor changes were made to cooling lines and cabling layouts.
%\textcolor{magenta}{[ADC help: Christoph, Dima, Becky]}

\begin{figure}[h!]
\begin{center}
\includegraphics[height=3.5in]{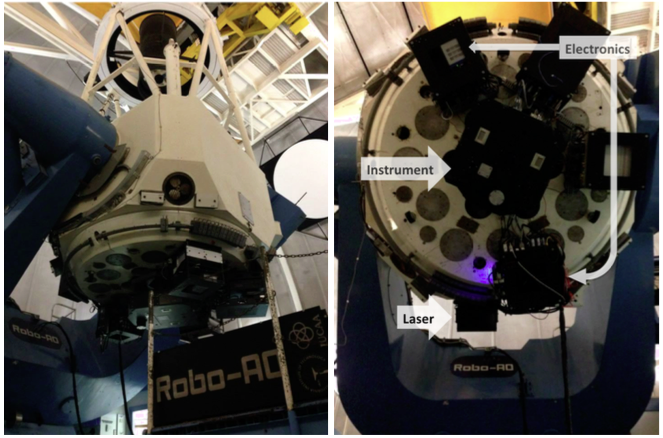}
\caption[fig:KP_setup]
{\label{fig:KP_setup} Images of the set-up of the Robo-AO Kitt Peak system on the 2.1-m telescope in Kitt Peak, Arizona. The instrument is attached on the bottom of the telescope and surrounded by the electronics. The laser projector is attached to the side of the telescope.}
%\vskip -0.25in
\end{center}
\end{figure}

\subsection{Software}

Unlike the robotic Palomar 1.5-m telescope, the 2.1-m is not yet automated, and the current computer interface is not robust enough to allow full automated control of the system by the Robo-AO instrument.  A manually operated observing system was developed; in this mode, the Robo-AO intelligent queue \cite{Riddle14} determines the next science target and presents the information to the operator, who manually points and centers the telescope, after which Robo-AO completes the rest of the observation automatically.  This has decreased Robo-AO efficiency by about 30\%; however, efforts are underway to automate the 2.1-m Telescope Control System (TCS) so that Robo-AO will again be able to operate automatically.  In spite of this decrease in efficiency and time spent optimizing the system, Robo-AO has completed over 14,000 science observations in the first six months of 2016. At Palomar, Robo-AO operated for a comparable amount of time in total, about 180 nights, and the total number of science observations was 20,000. We have also implemented software upgrades to improve the system, most notably the full automation of the laser and queue system interaction with the Laser Clearing House for satellite {avoidance\cite{Riddle14}}. %(http://adsabs.harvard.edu/abs/2014SPIE.9152E..1ER).

%%%%%%%%%%%%%%%%%%%%%%%%%%%%%%%%%%%%%%%%%%%%%%%%%%%%%%%%%%%%%
\section{Diagnostics Tools} 
\label{sec:Tools}

\subsection{Seeing}
\label{sec:seeing}
\begin{figure}[h]
\begin{center}
\includegraphics[width=3.5in]{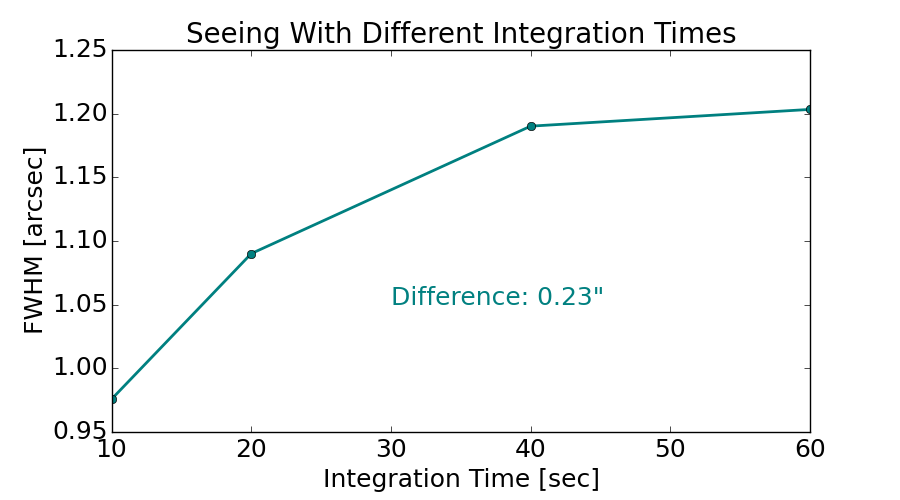}
\caption[fig:seeing_time]
{\label{fig:seeing_time} Seeing measurements corresponding to images of different integration times. The maximum difference in the measured seeing between the shortest integration time (10 seconds) and the longest (60 seconds) is $0.23''$.}
\end{center}
\end{figure}

By measuring atmospheric seeing we can estimate the current observing conditions. During laser acquisition, and before each target is observed, a ten second exposure time image of the brightest star in the field is taken with no AO correction. We measure the seeing by calculating the Full-Width Half-Maximum (FWHM) of the observed star's Point Spread Function (PSF) after fitting a 2D-Gaussian. %Typical integration times for science images are between 1 and 2 minutes, thus the 10 second integration time for the seeing image is significantly shorter. 
We did an experiment and have measured that the FWHM of a star's PSF increases with increasing exposure. Figure \ref{fig:seeing_time} shows the measured seeing corresponding to images of 10, 20, 40, and 60 second exposure times. We conclude that we are therefore underestimating the effective seeing of our science images by about 20$\%$. More tests need to be performed to understand whether the amount by which we appear to be underestimating the seeing is constant or varies with observing conditions. The limit of the asymptote seen in Figure \ref{fig:seeing_time} would be the real seeing.

\subsection{Strehl Ratios}
\label{sec:SR}
The Strehl ratio is the ratio between the peak of the observed PSF and the peak of a perfect PSF and is used as a diagnostics tool to determine image quality. Using the Mar\'echal approximation we can back out the root mean square (RMS) wavefront error from the Strehl ratio. As the Strehl ratio approaches 100$\%$ the image quality increases and approaches that of a perfect PSF. We used the measured Strehl ratio as metric while performing the optimizations to determine which settings yielded the highest Strehl ratios and therefore were the optimal settings. The Strehl ratio calculation is comprised of three main steps: 1) generating model PSFs for each observing band, 2) calculating the Strehl factor of the model PSF, and 3) dividing the normalized peak in the observed PSF by the Strehl factor to measure the Strehl ratio.

\subsubsection{Generating a model PSF}
\begin{figure}[h]
\begin{center}
\includegraphics[height=2in]{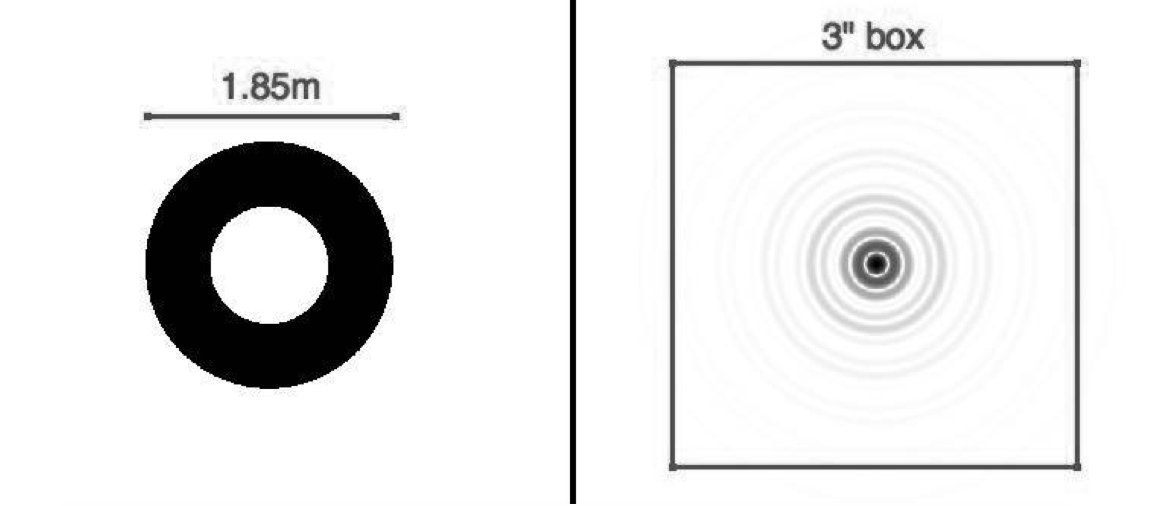}
%\subfloat{\includegraphics[height=2in]{pupil_plane.png}}
%\subfloat{\includegraphics[height=2in]{focal_plane.png}}
\caption[fig:pupil]
{\label{fig:pupil} \textit{Left:} Pupil plane image showing the primary mirror with a central obstruction due to the secondary mirror. The primary mirror has a diameter of 1.85-m and the secondary mirror of 0.879-m. \textit{Right:} Focal plane image showing the Airy diffraction pattern. Note that the first Airy ring has more intensity due to the central obstruction in the pupil image.}
%\vskip -0.25in
\end{center}
\end{figure}

The first step in generating a model PSF for an observing band is to generate a monochromatic PSF. This is done by starting with an oversampled image of the pupil (the primary mirror with the secondary mirror central obstruction), performing a Fourier Transform, and taking the square of the complex amplitudes to generate the intensity focal plane image. Figure \ref{fig:pupil} shows the oversampled pupil plane and focal plane images. The number of pixels constituting the image array ($N_{pix}$), as well as the number of pixels across the pupil ($D_{pix}$), are determined from relations between the wavelength ($\lambda$), diameter of the primary mirror ($D$), plate scale, and the desired oversampling factor ($f_{oversample}$). This relation is shown in the following equation:
\begin{equation}
    \frac{D_{pix}}{N_{pix}} \times \frac{\lambda\ (m)}{D\ (m)} \times 206265\ (''/rad)= \frac{plate\ scale\ (''/pix)}{f_{oversample}}
\end{equation}

%\begin{figure}[h]
%\begin{center}
%\includegraphics[scale=1,width=0.5\textwidth]{focal.png}
%\caption[fig:focal]
%{\label{fig:focal} Focal plane image showing the Airy diffraction pattern. Note that the first Airy ring has more intensity due to the central obstruction in the pupil image.}
%\vskip -0.25in
%\end{center}
%\end{figure}

\begin{figure}[h]
\begin{center}
\includegraphics[width=4.5in]{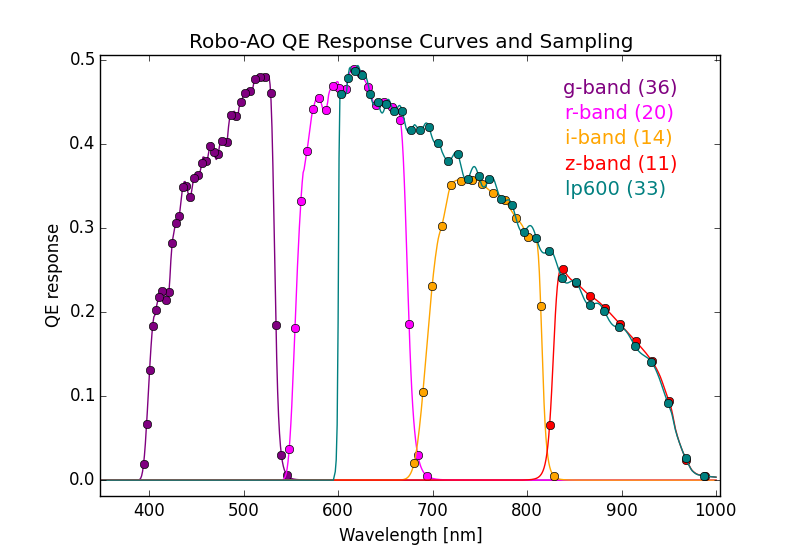}
\caption[fig:QE]
{\label{fig:QE} QE response curves over the different observing bands for Robo-AO\cite{Baranec14}. The sampling used to generate the monochromatic model PSFs is also shown and was generated using stepsizes constant in energy space (0.025 eV). The numbers in parantheses are the number of samples for each band.}
\end{center}
\end{figure}

The next step is to combine the monochromatic model PSFs in order to generate a model PSF for an entire observing band. This is done by normalizing each of the monochromatic PSFs by the sum of the energy in a 3$''$ box centered on the peak. The size of the box was chosen to include 95-99$\%$ of the energy for wavelengths corresponding to the 400nm-1000nm range. Using the measured Quantum Efficiency (QE) curves shown in Figure \ref{fig:QE} (from Baranec et al. (2014)\cite{Baranec14}), for each band, the normalized monochromatic PSFs are weighted then summed. For each observing band, monochromatic PSFs were generated at wavelengths corresponding to an even sampling in energy space, with increments of 0.025eV. %See Table \ref{tab:bands} for a summary of the sampling for each observing band in Robo-AO. 

\iffalse
\begin{table}[h]
%\caption{Observing bands for Robo-AO and sampling used to generate the model PSFs.} 
\label{tab:bands}
\begin{center}       
\begin{tabular}{|c|c|c|} %% this creates two columns
%% |l|l| to left justify each column entry
%% |c|c| to center each column entry
%% use of \rule[]{}{} below opens up each row
\hline
\rule[-1ex]{0pt}{3.5ex}  \textbf{Observing band} & \textbf{Wavelength range} & \textbf{Samples} \\
\hline
\rule[-1ex]{0pt}{3.5ex}  g-band & 400nm - 545nm & 36   \\
\hline
\rule[-1ex]{0pt}{3.5ex}  r-band & 545nm - 695nm & 20  \\
\hline
\rule[-1ex]{0pt}{3.5ex}  i-band & 675nm - 830nm & 14  \\
\hline
\rule[-1ex]{0pt}{3.5ex}  z-band & 815nm - 990nm & 11  \\
\hline
\rule[-1ex]{0pt}{3.5ex}  lp600 & 600nm - 990nm & 33  \\
\hline
%\rule[-1ex]{0pt}{3.5ex}  Y-band &  &   \\
%\hline
%\rule[-1ex]{0pt}{3.5ex}  H-band &  &   \\
%\hline
%\rule[-1ex]{0pt}{3.5ex}  J-band &  &  \\
%\hline
\end{tabular}
\end{center}
\caption{Observing bands for Robo-AO and number of samples in each band used to generate the model PSFs. The step-size in the sampling is uniform in energy space and corresponds to 0.025eV.} 
\end{table} 
\fi

\begin{figure}[h!]
    \centering
    \includegraphics[width=6.75in,height=1.35in]{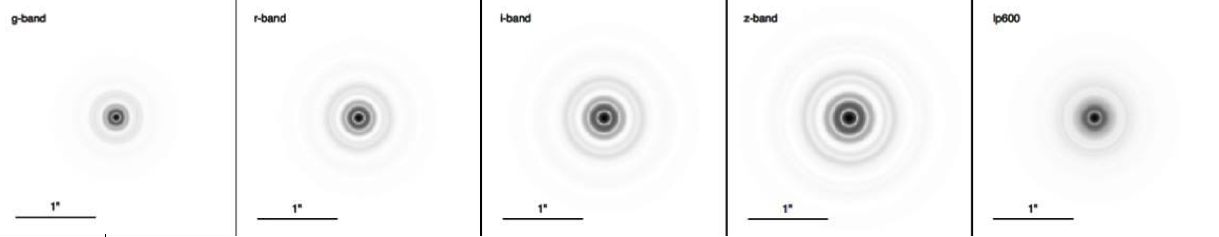}
    \caption{Model PSF images for the g, r, i, z, and lp600 bands. Each image is a $3'' \times 3''$ box and is scaled to the Robo-AO plate scale. As the wavelengths increase, the size of the Airy pattern also increases. Note that the lp600 filter includes a much larger range of wavelengths therefore the PSF is more blurred than for the other filters. }
    \label{fig:modelPSFs}
\end{figure}

The final step is to shift the center of the summed oversampled model PSF for each band to 9 grid positions within the size of a Robo-AO pixel, then scale each of these shifted images to the Robo-AO plate scale. The purpose of generating these sub-Robo-AO pixel shifted centers is to account for the fact that the peak may not land in the center of a pixel due to the random location of a star during observations. Next, for each shifted and scaled image, the Strehl factor is calculated. This is done by dividing the intensity peak by the sum of the intensities in the same 3$''$ square box. Finally, the Strehl factor for a given band is the mean of the Strehl factors for each shifted PSF in order to generate the Strehl factor for the average of star positions within the pixel. The model PSFs for each of the bands in the optical, observable with Robo-AO, are shown in Figure \ref{fig:modelPSFs}. Each model PSF is scaled to the Robo-AO plate scale and is a 3$''$ box.

\subsubsection{Calculating the Strehl ratio}
Finally, to calculate the Strehl ratio of an observed image, the peak intensity is divided by the sum of the energy in a 3$''$ box, then this normalized peak is divided by the Strehl factor to yield the Strehl ratio. 

% commented out these multiple lines

\subsection{Contrast Curves}
\label{sec:contrast}
The Robo-AO data reduction pipeline described in Law et al {(2009)\cite{Law09}}, (2014)\cite{Law14} performs flat field calibration, dark subtraction, and post-facto shift and add on the individual frames in a given observation to produce the final science frame. A new pipeline has been developed to quickly produce PSF-subtracted frames and contrast curves from these science images (Jensen-Clem et al. 2016, in prep) and are available the next day. After identifying the target star in the final frame, a $2.8'' \times 2.8''$ cutout around the star is spatially filtered to reduce the contribution of the uncorrected stellar halo. This cutout is cross correlated with cutouts of all previous stars observed with Robo-AO at Kitt Peak. The five stars with the highest correlation become the PSF library with which a reference PSF is constructed and subtracted using the KLIP algorithm. The contrast curves are generated by calculating the standard deviations of resolution element sized aperture sums at given distances from the central star. The Vortex Image Processing pipeline (VIP, Gomez Gonzalez et al. 2016, in prep) is used to implement KLIP and generate contrast curves. Figure \ref{fig:contrasts} below shows representative contrast curves for the first six months of observations at Kitt Peak.

\begin{figure}[h!]
\begin{center}
\includegraphics[height=2.4in]{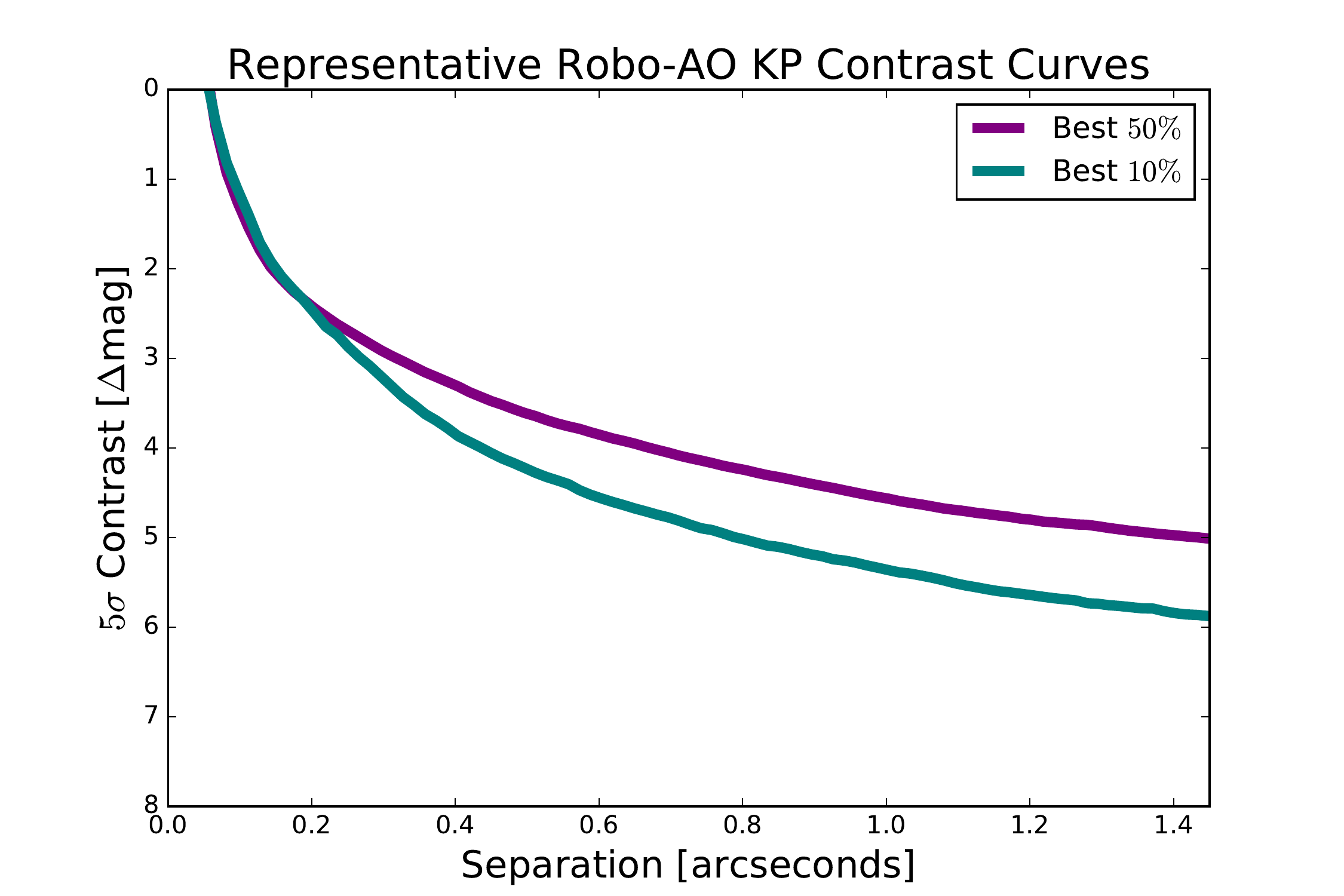}%scale=1,width=0.5\textwidth]{representative_contrast_curve.pdf}
\caption[fig:contrasts]
{\label{fig:contrasts} The curves representing the $50\%$ and $10\%$ best contrast at 1.4" are plotted.}
\end{center}
\end{figure}

\subsection{Nightly Summary Report Website}
\label{sec:website}

\begin{figure}[h]
\begin{center}
\includegraphics[width=6.7in]{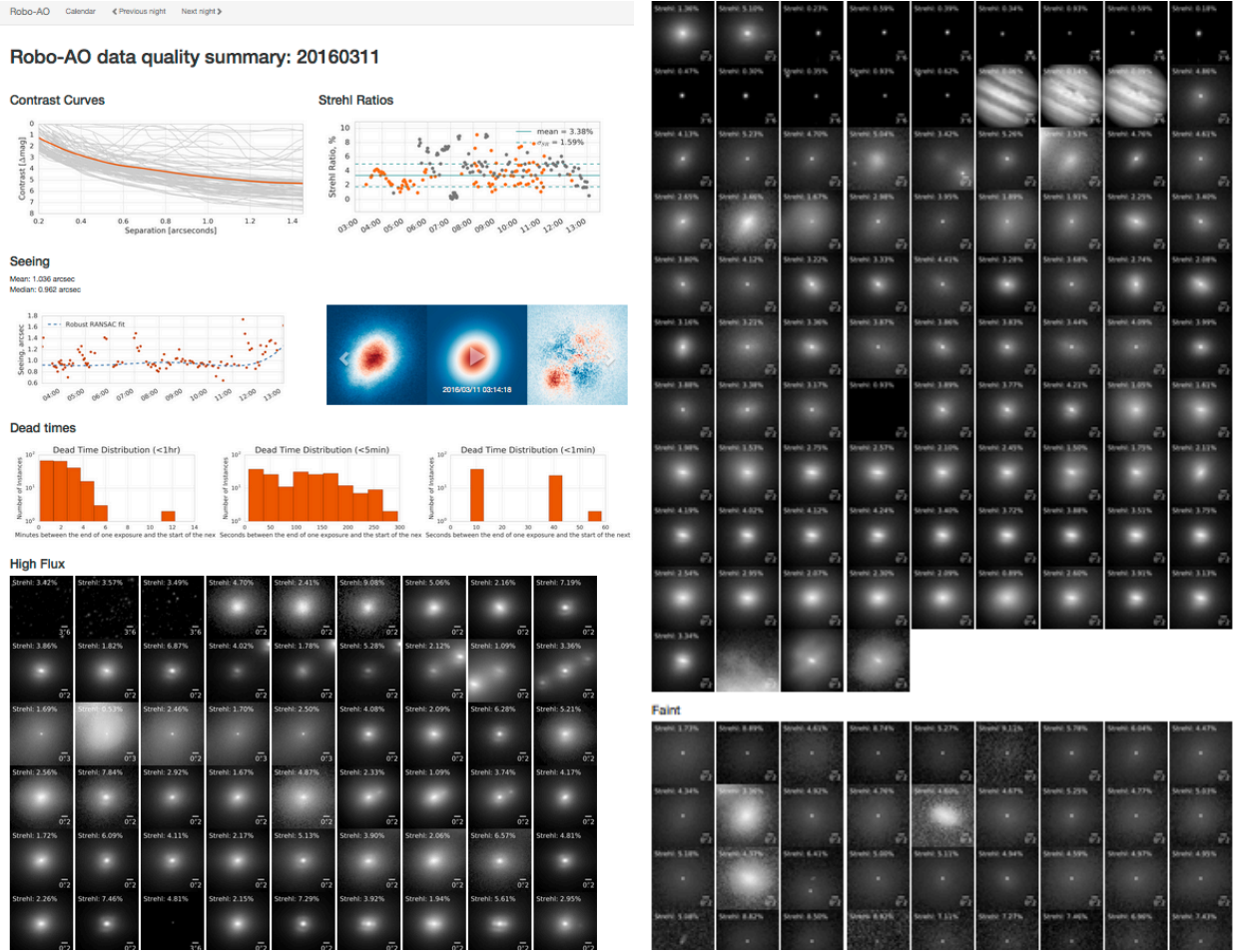}
\caption[fig:web]
{\label{fig:web} Screenshot of the nightly summary report website, which shows each night's observing conditions and performance.}
\end{center}
\end{figure}

A nightly summary report website is generated after each night of observing and is available the next day along with all of the past observing nights. This website contains information about the observing conditions and performance of the instrument. The contrast curves, Strehl Ratios, and seeing measurements are plotted for each night and all observed images are displayed. Each image can be clicked on so full sized and cropped images are shown along with the Strehl Ratio, contrast curve, and full header information. Figure \ref{fig:web} is a screenshot of the website and shows its layout.

%%%%%%%%%%%%%%%%%%%%%%%%%%%%%%%%%%%%%%%%%%%%%%%%%%%%%%%%%%%%%
\section{Optimizations}
\label{sec:Opt} 

We performed a series of tests in 2016 May 17-20 in order to optimize the system at Kitt Peak. After each test was performed, the reduced data was available the next day. The measured Strehl ratios were used to determine the highest quality images and thus the optimal settings for the instrument.

\subsection{Quad-Cell Response Look-Up Tables} 
\label{sec:lintables}

\begin{figure}[h]
\begin{center}
\includegraphics[scale=1,width=0.5\textwidth]{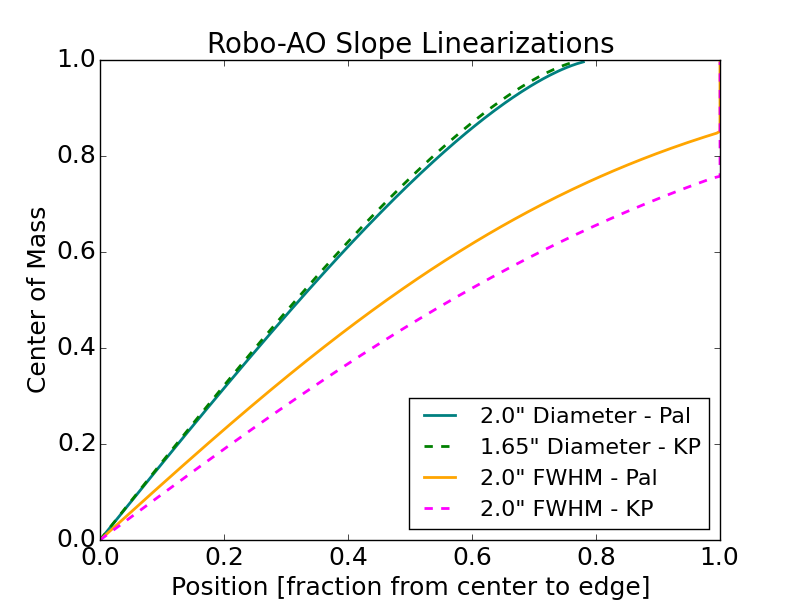}
\caption[fig:slopelin]
{\label{fig:slopelin} Quad-cell response look-up-tables, for calibrations (circular) and on-sky observations (gaussian). Matches the PSF position on the wavefront sensor quad-cell with the center-of-mass algorithm output to linearize the response for different sources. This plot shows the different response tables for Palomar and Kitt Peak.}
%\vskip -0.25in
\end{center}
\end{figure}

The quad-cell response look-up tables are used to convert between the measurement of the center of mass of the intensity on the Shack-Hartmann wavefront sensor images and the corresponding physical position on the wavefront sensor in order to apply the appropriate corrections on the deformable mirror. Due to the different plate scale between the Palomar and Kitt Peak systems, the look-up tables needed to be regenerated. The light from the telescope simulator is circular while the light from the on-sky laser is roughly gaussian, so the response curves are different. Look-up tables were produced for both types of sources and are shown in Figure \ref{fig:slopelin}, also comparing the Palomar and Kitt Peak systems. 

\subsection{Laser Range Gate} 
\label{sec:laser}
\begin{figure}[h]
%\vskip -0.25in
\begin{center}
\includegraphics[width=6.6in]{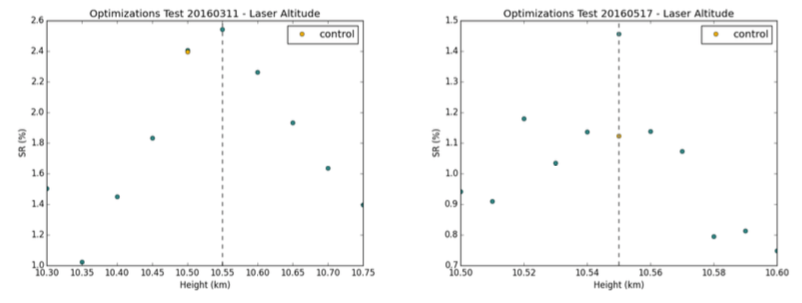}
\caption[fig:LGS_alt_plot]
{\label{fig:LGS_alt_plot} Plot of the Strehl ratios corresponding to the laser set to various distances along the line of sight. During the optimizations test on 2016 March 11 (\textit{left}) we used step sizes of 0.05km and found a clear peak at 10.55km. During a later optimizations run, on 2016 May 17 (\textit{right}), we used smaller step sizes of 0.01km around 10.55km. Again there is a clear peak at 10.55km, which is now the current altitude setting of the instrument. Note that although the control image has a lower Strehl ratio at 10.55km, this image was taken before stepping through the different altitudes and the seeing was worse at that time and improved as the test went on. Therefore, the Strehl ratio being greater at 10.55km than at 10.54km and 10.56km is significant because the seeing was similar.}
\end{center}
\end{figure}

The laser range gate controls two aspects of the laser guide star: 1) the focus distance along the line of sight in the atmosphere and 2) depth measurement, the size of the slice of atmosphere we are collecting the light from. They are respectively controlled by the delay in opening the laser shutter and the length of time the shutter remains open. The laser is a pulsed Rayleigh beam, so it backscatters at every altitude in the atmosphere. We want to collect the backscatter that corresponds to the layer of atmosphere at which the laser is focused ($\approx$10km), since the laser is traveling at the speed of light, the round trip corresponds to a delay of about 66$\mu$s. For our first set of optimizations tests (2016 March 11) we set the delay to correspond to focus distances ranging from 10.3km to 10.75km with stepsizes of 0.05km. While we ran the tests on the focus distance, we kept the depth constant at 450m, so this corresponds to the shutter being open for 3$\mu$s. We found a clear improvement in image quality (measured by a higher Strehl Ratio, explained in $\S$\ref{sec:SR}) at an altitude of 10.55km. We then ran the optimizations again on 2016 May 17, this time stepping through values around 10.55km by increments of 0.01km and again confirmed that the focus distance set at 10.55km yielded the best image quality. Figure \ref{fig:LGS_alt_plot} shows the plots of the Strehl Ratios corresponding to different laser focus distances. 

%\begin{figure}[h]
%\begin{center}
%\includegraphics[width=6in]{20160517_LGS_alt_images.png}
%\caption[fig:LGS_alt_img]
%{\label{fig:LGS_alt_img} Images of of the same star taken with the laser altitude changing from 10.5km to 10.6km. The Strehl ratio is shown in the top right corner. These images were taken during optimizations test run on May 17$^{th}$ 2016.}
%\vskip -0.25in
%\end{center}
%\end{figure}

\begin{figure}[h]
\begin{center}
\includegraphics[width=3.3in]{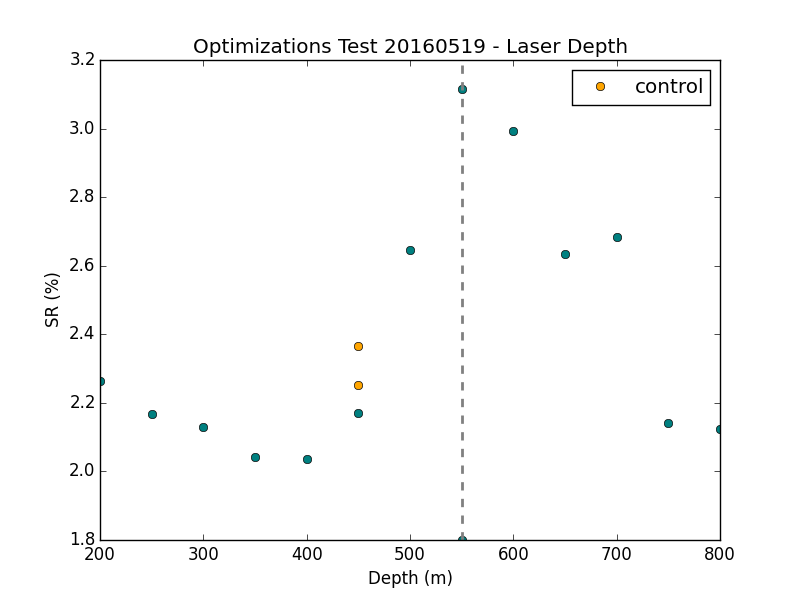}
%\subfloat{\includegraphics[width=3.3in]{OPT_20160519_LGS_depth.png}}
%\subfloat{\includegraphics[width=3.3in]{20160519_LGS_depth_images.png}}
\caption[fig:LGS_depth_plot]
{\label{fig:LGS_depth_plot} Plot of the Strehl ratios corresponding to the laser set to various depth measurements. There is a clear peak at 550m, which is now the current range gate setting of the instrument.}
%\vskip -0.25in
\end{center}
\end{figure}

There are two main considerations to take into account when setting the depth of laser measurement. A longer range gate will allow us to collect more light but will also cause image elongations on the wavefront sensor as the light comes from distances farther away from the focus. During the following test, on 2016 May 19, with the laser now set to an altitude of 10.55km, we stepped through a range of depths, from 200m to 800m by increments of 50m and again found a clear peak in the Strehl ratios at 550m, as seen in Figure \ref{fig:LGS_depth_plot}. The current laser range gate settings thus correspond to 10.55km and 550m.

%\begin{figure}[h]
%\begin{center}
%\includegraphics[width=6in]{20160519_LGS_depth_images.png}
%\caption[fig:LGS_depth_img]
%{\label{fig:LGS_depth_img} Plot of the Strehl ratios corresponding to the laser set to various altitudes. There is a clear peak at 10.55km, which is now the current altitude setting of the instrument.}
%\vskip -0.25in
%\end{center}
%\end{figure}

\subsection{AO Loop Gain Settings} 
\label{sec:gain}
\begin{figure}[h!]
\begin{center}
\includegraphics[width=6in]{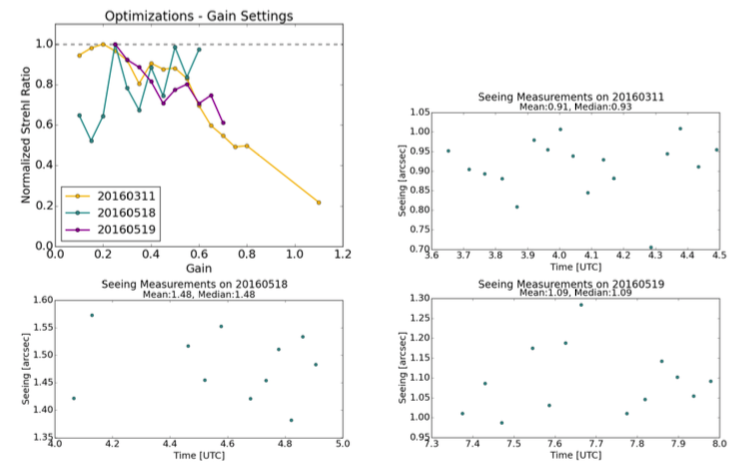}
%\subfloat{\includegraphics[width=3in]{OPT_alldates_gains.png}}
%\subfloat{\includegraphics[width=3in]{seeing_20160311_gain.png}}
%\subfloat{\includegraphics[width=3in]{seeing_20160518_gain.png}}
%\subfloat{\includegraphics[width=3in]{seeing_20160519_gain.png}}
\caption[fig:gains_plot]
{\label{fig:gains_plot} Plot of the Strehl ratios corresponding to images corrected with different gain settings. The test was repeated on several nights (2016 March 11 and 2016 May 18 $\&$ 19) so the Strehl ratios are each normalized by the maximum Strehl ratio of each respective night due to different seeing conditions on different nights. Also shown are the seeing conditions during each of the nights. The time is shown in decimal UTC time.}
%\vskip -0.25in
\end{center}
\end{figure}

The AO loop gain setting controls how much of a correction to the distorted wavefront should be applied with the deformable mirror based on the measured error from the wavefront sensor. Since the atmosphere is constantly changing, there will always be a delay between the time the wavefront distortion is measured on the wavefront sensor and the instructions to correct it are received and applied by the deformable mirror. If the full correction were to be applied, we would risk over-correcting the wavefront and introducing larger distortions instead, causing the loop to go unstable. The gain setting is a value usually between 0 and 1 and represents the fraction of the correction to be applied. During the nights of 2016 March 11 and 2016 May 18 - 20 we ran several tests while changing the gain settings. The results were not as clear as with the laser tests. Figure \ref{fig:gains_plot} shows the results of the gain settings tests. The original gain setting was 0.36, which is our predicted optimal gain setting for Kitt Peak. This number was obtained by converting the optimal gain setting from Palomar using the factor difference between the plate scales and the new quad-cell response look-up tables (see $\S$\ref{sec:lintables}). The results of the tests were not conclusive enough to justify changing it from that value. 

\subsection{Spherical Slope Offsets} 
\label{sec:sph}
\begin{figure}[h]
\begin{center}
\includegraphics[width=6in]{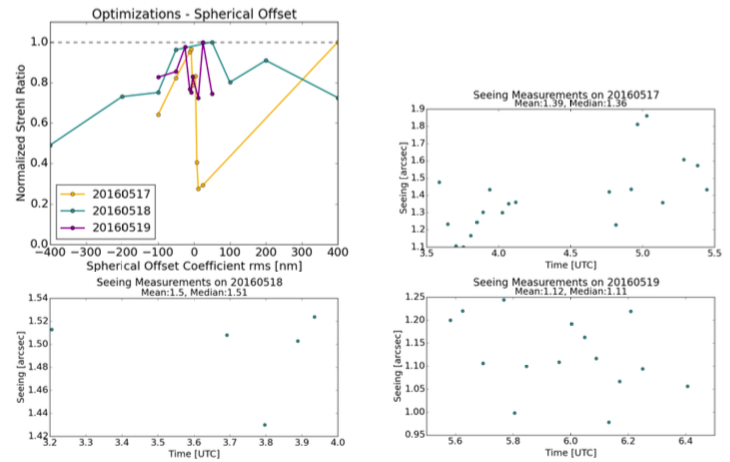}
%\subfloat{\includegraphics[width=3in]{OPT_alldates_sph.png}}
%\subfloat{\includegraphics[width=3in]{seeing_20160517_sph.png}}
%\subfloat{\includegraphics[width=3in]{seeing_20160518_sph.png}}
%\subfloat{\includegraphics[width=3in]{seeing_20160519_sph.png}}
\caption[fig:sph_plot]
{\label{fig:sph_plot} Plot of the Strehl ratios corresponding to images corrected with different spherical slope offsets. The test was repeated on several nights (2016 May 17, 18 \& 19) so the Strehl ratios are each normalized by the maximum Strehl ratio of each respective night due to different seeing conditions on different nights. Also shown are the seeing conditions during each of the nights. The time is shown in decimal UTC time.}
%\vskip -0.25in
\end{center}
\end{figure}

Due to the addition of an aspheric condenser lens, we wanted to confirm that no significant non-common path spherical aberrations were introduced. We thus performed a spherical slope offset test by adding slope offsets corresponding to different amplitudes of Zernike spherical aberration. We performed this test on the nights of 2016 May 17-19 and the resulting normalized Strehl Ratios are shown in Figure \ref{fig:sph_plot} along with the seeing conditions during the times of the tests for each night. Note that the point corresponding to a coefficient of 400nm on 2016 May 17 was taken at 3.7AM during an improvement in seeing ($1.1''$ compared to the mean of $1.39''$) and its high Strehl ratio is thus not reliable. We concluded that no significant non-common path spherical aberrations had been introduced because the higher Strehl ratios corresponded to the coefficients of small values surrounding zero. We are thus currently not adding slope offsets.

%%%%%%%%%%%%%%%%%%%%%%%%%%%%%%%%%%%%%%%%%%%%%%%%%%%%%%%%%%%%%%%%%%%
\section{Performance Analysis and Comparison to Palomar} 
\label{sec:Perf}

\subsection{Seeing Comparison}
\label{sec:seeing_comp}

\begin{figure}[h!]
\begin{center}
\includegraphics[scale=1,width=0.5\textwidth]{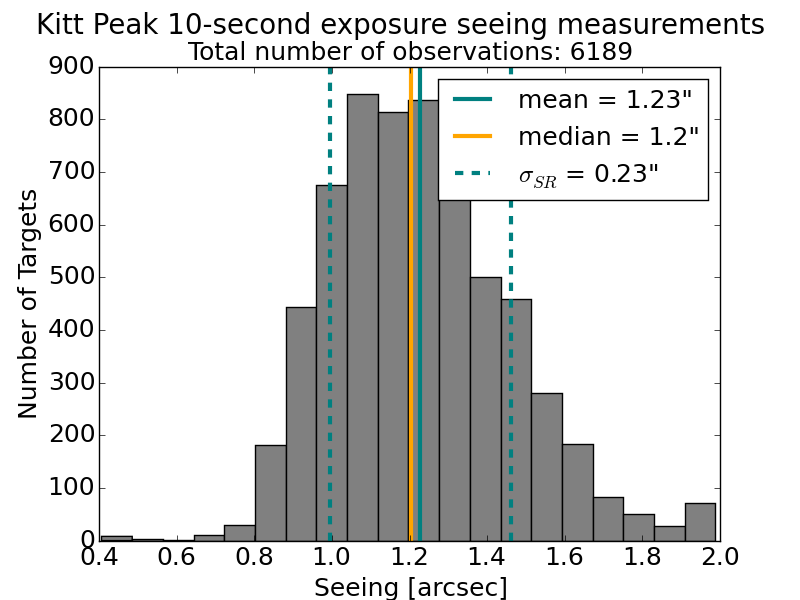}
\caption[fig:seeing_KP]
{\label{fig:seeing_KP} Histogram of the seeing measurements at Kitt Peak from 10-second exposure time images. This includes all of the observations made in all bands between the dates of 2015 November 14 and 2016 June 11. The mean and standard deviation are also shown.}
\end{center}
\end{figure}

In order to analyze the performance of Robo-AO at Kitt Peak and compare it to its previous performance at Palomar, the observing conditions need to be taken into account. The median seeing for stellar targets observed in the i'-band at Palomar was reported to be $1.1''$ using 20 second exposure time seeing {measurements\cite{Baranec14}}. This value is in rough agreement with the R-band median seeing of $1.1''$ reported in Cenko et al. {(2006)\cite{Cenko06}}. After running a similar analysis at Kitt Peak, and considering all targets, all observing bands (most observations are done in the i'-band), %(although 56$\%$ of observations are in i'-band) 
and only 10 second exposure measurements (see $\S$\ref{sec:seeing}), we found the median seeing to be $1.2''$, which is likely to be underestimating the true seeing. This is significantly higher than the median seeing reported by the 4-m telescope at Kitt Peak\cite{Code98} of $0.8''$. Figure \ref{fig:seeing_KP} shows the histogram of the 10-sec exposure time seeing measurements at Kitt Peak. We are currently looking into ways of improving the dome seeing. KPNO staff used a thermal camera to identify heat sources in the dome and generated a report with several recommendations on how to address these thermalization problems. Using these recommendations we have created a prioritized list of tasks to pursue to work on improving the dome seeing.

\subsection{Strehl Ratio Comparison}
\label{sec:strehl_comp}

After calculating the Strehl ratios for all of the observations done at Palomar and those done up to 2016 June 11 at Kitt Peak, we can compare the data quality between the two systems. The observations included in this analysis are those that passed the criteria to be considered `good' observations. Law et al. (2014)\cite{Law14} found that core size was the best indicator of performance and image quality. Images with core sizes smaller than the diffraction limit were instead due to the stacking on a noise spike because of the low signal-to-noise ratio (SNR). Thus, for this study, we excluded images with core sizes smaller than 0.14$''$. Extended sources such as planets, moons, and asteroids were also excluded from the calculations as the model PSF assumed point sources. Targets with companions were not specifically excluded from the calculation, which could contribute to slightly lower Strehl values. 

\begin{figure}[h!]
\begin{center}
\includegraphics[scale=1,width=0.5\textwidth]{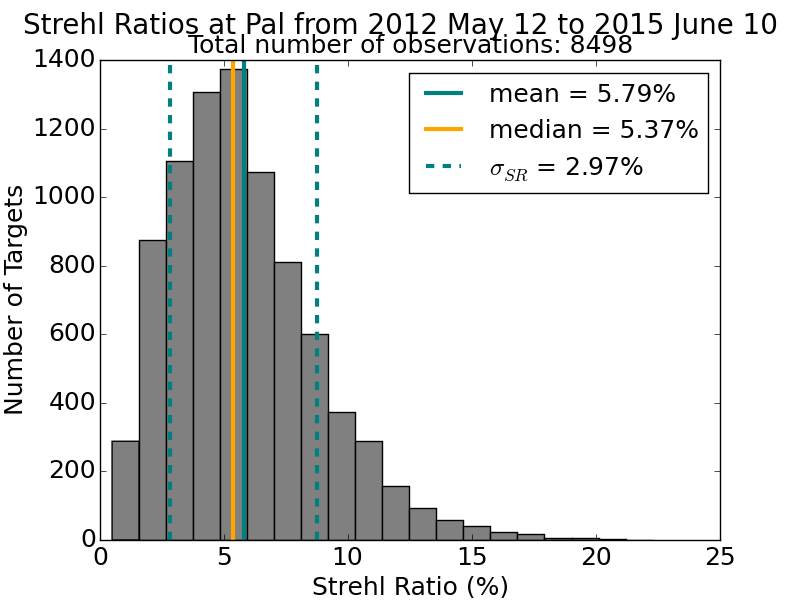}
\caption[fig:SR_Pal]
{\label{fig:SR_Pal} Histogram of all of the Strehl ratios from observations included in this analysis from 2012 May 12 to 2015 June 10 at Palomar. The mean and standard deviation are also shown.}
\end{center}
\end{figure}

\begin{figure}[h!]
\begin{center}
\includegraphics[scale=1,width=\textwidth]{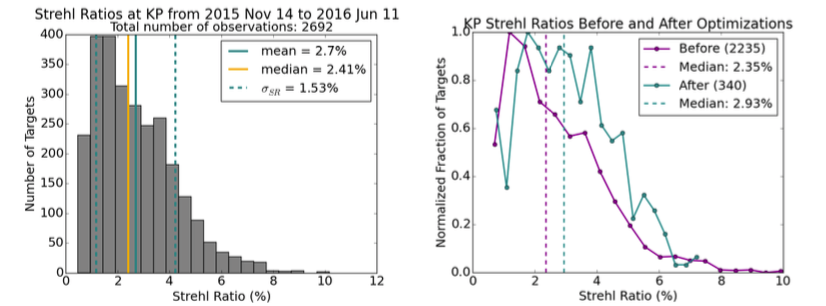}
%\subfloat{\includegraphics[scale=1,width=0.48\textwidth]{SR_KP_all.png}}
%\subfloat{\includegraphics[scale=1,width=0.48\textwidth]{SR_KP_pre-post_opts.png}}
\caption[fig:SR_KP]
{\label{fig:SR_KP} \textit{Left:} Histogram of all of the Strehl ratios from observations included in this analysis from 2015 November to 2016 June at Kitt Peak. The mean and standard deviation are also shown. \textit{Right:} Comparison of the distribution of Strehl ratios before and after Robo-AO was optimized in May 2016.}
\end{center}
\end{figure}

The Strehl ratios measured for observations during Palomar operations (2012 May 12 to 2015 June 10) and that passed the selection criteria described above are shown in a histogram in Figure \ref{fig:SR_Pal}. The median Strehl ratio was found to be 5.37$\%$ for a total of 8498 observations. Figure \ref{fig:SR_KP} (left) shows the Strehl ratios for the observations from Kitt Peak, also using the same selection criteria, and observed between 2015 November 14 and 2016 June 11. The median Strehl ratio was found to be 2.41$\%$ for a total of 2692 observations, which is significantly lower than at Palomar. As we work on improving the dome seeing (see previous section, \S \ref{sec:seeing_comp}), we expect our Strehl ratios to improve. The optimizations described in \S \ref{sec:Opt} have also helped improve the image quality, as seen in Figure \ref{fig:SR_KP} (right) with an increase in median Strehl ratio when comparing observations from before the optimizations were done (2014 Nov to 2016 May), to after the optimizations were done (2016 May to 2016 June). The median Strehl ratio increased from 2.35$\%$ for 2235 observations to 2.93$\%$ for 340 observations. As we continue to optimize the system and improve dome seeing we also expect the Strehl ratios to increase.

%\begin{figure}[h]
%\begin{center}
%\includegraphics[scale=1,width=0.5\textwidth]{SR_KP_pre-post_opts.png}
%\caption[fig:SR_prepost]
%{\label{fig:SR_prepost} Comparison of Strehl Ratios before and after Robo-AO was optimized in May 2016.}
%\end{center}
%\end{figure}

%%%%%%%%%%%%%%%%%%%%%%%%%%%%%%%%%%%%%%%%%%%%%%%%%%%%%%%%%%%%%
\section{IR Camera} 
\label{sec:IRcam}
\begin{figure}[h]
\begin{center}
\includegraphics[width=4in]{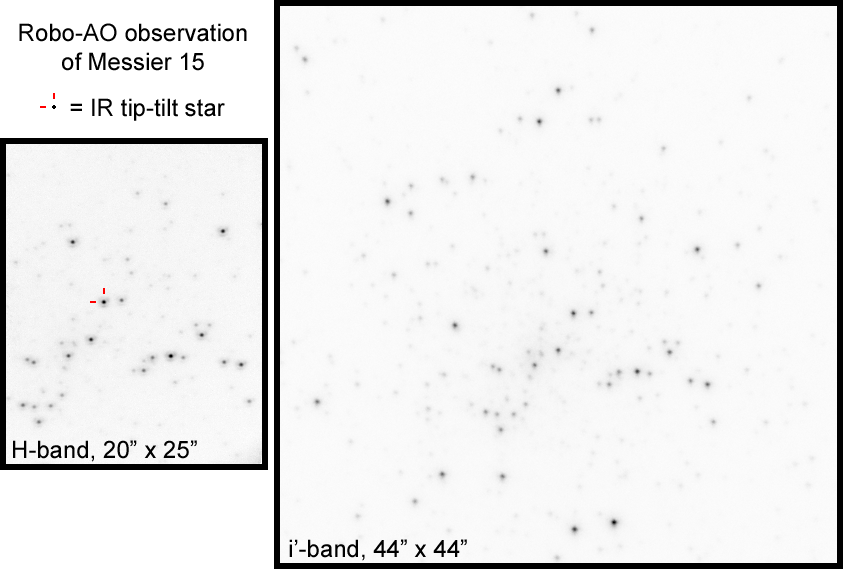}
\caption[fig:M15]
{\label{fig:M15} Palomar observations in H-band and i-band of M15 with infrared tip-tilt guiding. These images were simultaneously taken during demonstration tests in 2014 with the Robo-AO Palomar {system\cite{Baranec15}}.}
\end{center}
\end{figure}

\begin{figure}[h]
\begin{center}
\includegraphics[scale=1,width=0.5\textwidth]{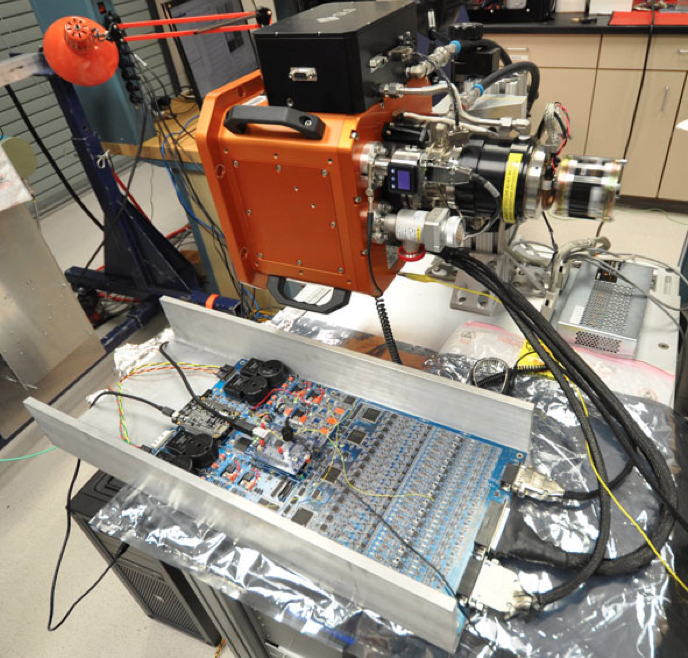}
\caption[fig:SAPHIRA]
{\label{fig:SAPHIRA} GLS Scientific cryostat (orange) with the SAPHIRA detector installed inside and the PB1 (``PizzaBox") electronics circuit board (blue) in the lab in Hilo, Hawai'i.}
\end{center}
\end{figure}

We are planning the addition of an infrared camera to Robo-AO KP, which will allow us to widen the scope of possible targets, specifically to much redder and low-mass objects, and to conduct the largest ever AO surveys in the infrared. The camera will double as an IR science detector and as a tip-tilt sensor that takes advantage of significant image sharpening in the infrared. The IR camera uses brand new infrared avalanche photodiode array technology (SAPHIRA {detector\cite{Finger14}}), which has so far only been used for fringe {tracking\cite{Eisenhauer16}}, demonstration tests{\cite{Atkinson16}$^,$\cite{Goebel16}}, and with the prototype Robo-AO system\cite{Baranec15} on a limited basis in 2014 (see Figure \ref{fig:M15}). This will be the first long-term deployment of such a camera on an adaptive optics system for science applications.  
%\textcolor{magenta}{[Help (software): Reed]}
The cryostat was delivered to the team in Hilo, Hawai'i in mid-June 2016 and will be in the lab in Hilo for the duration of summer 2016, undergoing testing and in order to finalize the electronics and software interface. The IfA has developed PB1 (``PizzaBox") electronics which can be used in different configurations for Hawaii-2RG, Hawaii-4RG, and SAPHIRA detectors. In our case, using the single board configuration for the 320$\times$256 SAPHIRA detector, it will have 34 readout channels and is capable of reaching a 3$\;$Mpixel/sec sampling rate per channel. This allows for a 1kHz full frame readout rate. The assembled camera is shown in the lab in Hilo in Figure \ref{fig:SAPHIRA}. Once the testing is completed, the IR detector and cryostat will then be taken to Kitt Peak, Arizona to be added to the Robo-AO KP system.

\subsection{Science Programs}
The first science program to be conducted with Robo-AO Kitt Peak in the infrared will be a survey of nearby young stars to search for brown dwarfs and wide-orbit exoplanets, this will be the largest ever adaptive optics survey of its kind. Brown dwarfs are objects that form as a result of not being massive enough to form stars but are also too massive to be considered planets ($< 13~M_{Jup}$) by the current IAU guidelines. They can therefore be seen as a link between stars and planets. Understanding how they form and evolve will provide insight on the formation and evolution of stars and planets as well, by understanding the differences in the processes. With each new substellar object (brown dwarf or exoplanet) discovered and characterized, we have the opportunity to gain insight on the diversity of such objects, their occurrence, and their evolution. There have been few wide-orbit exoplanets found around young stars. Similarly, only about 1$\%$ of known brown dwarfs are found as companions. Therefore, a large sample of targets is necessary to find these rare objects and understand their frequency. Robo-AO is the perfect instrument for such a large-scale survey because of its excellent angular resolution and automated efficiency it has the ability to observe large numbers of targets per night and will allow us to drastically increase our sample size relative to other studies and allow us to discover many more of these rarely detected substellar objects. %\textcolor{red}{[To do: add about brown dwarf binaries and use for isochrone models, also check that frequency is about 10$\%$]}

%%%%%%%%%%%%%%%%%%%%%%%%%%%%%%%%%%%%%%%%%%%%%%%%%%%%%%%%%%%%%
\section{Conclusion}
\label{sec:Concl}

\iffalse
Summarize results, and describe next steps:
\begin{itemize}
\item future optimizations calibrations
\item IR camera: on sky
\item Telescope automation
\item Open to wider astro community for 2017A through NOAO -- nevermind don't include
\end{itemize}
\fi 

Robo-AO was decommissioned from the Palomar 1.5-m telescope in California, reconfigured, and successfully commissioned on the 2.1-m telescope in Kitt Peak, Arizona, in less than 6 months during the summer and fall of 2015. Robo-AO is currently operating on sky and collecting science data while undergoing optimizations. Two science papers have been published using Robo-AO Kitt Peak data. Vanderburg et al. {(2016)\cite{Vanderburg16a}}$^,$\cite{Vanderburg16b} used Robo-AO to image stars with planet candidates and ensure there were no contaminants. Several optimizations tests were conducted during the spring of 2016, including regenerating the quad-cell response look-up tables, finding the optimal focus distance along the line of sight and range gate for the laser guide star system, testing different gain settings, and spherical slope offsets. These tests have led to an increase in the Strehl ratios of the images collected since then. The seeing measurements have been higher than they were at Palomar and than what is reported as the median seeing by the 4-m telescope at Kitt Peak. We are working on improving the dome seeing by running thermalization analyses with the help and support of NOAO. Improving the dome seeing conditions will help improve the image quality. We are in the process of partially automating the Kitt Peak 2.1-m telescope. The addition of an infrared camera will allow us to simultaneously image in the visible and infrared, as well as conducting tip-tilt sensing in the infrared and widen our sky coverage. This will be the first such deployment of a SAPHIRA detector specifically for science imaging and will lead to the largest AO survey in the infrared.

%%%%%%%%%%%%%%%%%%%%%%%%%%%%%%%%%%%%%%%%%%%%%%%%%%%%%%%%%%%%%
\acknowledgments      
 Robo-AO KP is a partnership between the California Institute of Technology, University of Hawaii Manoa, University of North Carolona, Chapel Hill, the Inter-University Centre for Astronomy and Astrophysics, and the National Central University, Taiwan. Robo-AO KP was supported by a grant from Sudha Murty, Narayan Murthy, and Rohan Murty. The Robo-AO instrument was developed with support from the National Science Foundation under grants AST-0906060, AST-0960343, and AST-1207891, the Mt.$\;$Cuba Astronomical Foundation, and by a gift from Samuel Oschin. C.B. acknowledges support from the Alfred P. Sloan Foundation and R.J.C. from the National Science Foundation Graduate Research Fellowship under grant No. DGE-1144469.
%%%%%%%%%%%%%%%%%%%%%%%%%%%%%%%%%%%%%%%%%%%%%%%%%%%%%%%%%%%%%
%%%%% References %%%%%

\bibliography{Salama_SPIE2016_Proc.bib}   %>>>> bibliography data in report.bib

\begin{thebibliography}{10}

\bibitem{Baranec14}
Baranec, C., Riddle, R., Law, N.~M., et~al., ``High-efficiency autonomous laser
  adaptive optics,'' {\em ApJ Letters}~{\bf 790},  L8 (2014).

\bibitem{JOVE13}
Baranec, C., Riddle, R., Law, N.~M., et~al., ``Bringing the visible universe
  into focus with {R}obo-{AO},'' {\em Journal of Visualized Experiments}~{\bf
  72},  e50021 (2013).

\bibitem{Atkinson14}
Atkinson, D., Hall, D., Baranec, C., et~al., ``Observatory deployment and
  characterization of {SAPHIRA} {H}g{C}d{T}e {APD} arrays,'' in [{\em High
  Energy, Optical, and Infrared Detectors for Astronomy VI,
  915419}{\nolinebreak\hspace{0.1em}]},  {\em Proc. SPIE} {\bf 9154},
  915419--12 (2014).

\bibitem{Baranec15}
Baranec, C., Atkinson, D., Riddle, R., et~al., ``High-speed imaging and
  wavefront sensing with an infrared avalanche photodiode array,'' {\em
  ApJ}~{\bf 809},  70 (2015).

\bibitem{Atkinson16}
Atkinson, D.~E., Hall, D. N.~B., Baker, I.~M., et~al., ``Next-generation
  performance of {SAPHIRA} {H}g{C}d{T}e {APD}s,'' in [{\em High Energy,
  Optical, and Infrared Detectors for Astronomy VII
  9915E}{\nolinebreak\hspace{0.1em}]},  {\em Proc. SPIE} {\bf 9915},  9915--22
  (2016).

\bibitem{Riddle14}
Riddle, R.~L., Hogstrom, K., Papadopoulos, A., Baranec, C., and Law, N.~M.,
  ``The {R}obo-{AO} automated intelligent queue system,'' in [{\em Software and
  Cyberinfrastructure for Astronomy III, 91521E}{\nolinebreak\hspace{0.1em}]},
  {\em Proc. SPIE} {\bf 9152},  91521E--13 (2014).

\bibitem{Law09}
Law, N.~M., Mackay, C.~D., Dekany, R.~G., et~al., ``Getting {L}ucky with
  adaptive optics: Fast adaptive optics image selection in the visible with a
  large telescope,'' {\em ApJ}~{\bf 692}(1),  924--930 (2009).

\bibitem{Law14}
Law, N.~M., Morton, T., Baranec, C., et~al., ``Robotic laser adaptive optics
  imaging of 715 {K}epler exoplanet candidates using {R}obo-{AO},'' {\em
  ApJ}~{\bf 791}(1),  35 (2014).

\bibitem{Cenko06}
Cenko, S.~B., Fox, D.~B., Moon, D.-S., et~al., ``The automated {P}alomar 60
  inch telescope,'' {\em PASP}~{\bf 118},  1396--1406 (2006).

\bibitem{Code98}
Code, A.~D., Claver, C.~F., Goble, L.~W., Jacoby, G.~H., and Sawyer, D.~G.,
  ``{WIYN} active optics: a platform for {AO},'' in [{\em Software and
  Cyberinfrastructure for Astronomy III, 91521E}{\nolinebreak\hspace{0.1em}]},
  {\em Proc. SPIE} {\bf 3353},  649--657 (1998).

\bibitem{Finger14}
Finger, G., Baker, I., Alvarez, D., et~al., ``{SAPHIRA} detector for infrared
  wavefront sensing,'' in [{\em Adaptive Optics Systems IV
  9148E}{\nolinebreak\hspace{0.1em}]},  {\em Proc. SPIE} {\bf 9148},  9148--17
  (2014).

\bibitem{Eisenhauer16}
Eisenhauer, F., Perrin, G.~S., Henning, T., et~al., ``First light for
  {GRAVITY},'' in [{\em Optical and Infrared Interferometry and Imaging V
  9907E}{\nolinebreak\hspace{0.1em}]},  {\em Proc. SPIE} {\bf 9907},  9907--6
  (2016).

\bibitem{Goebel16}
Goebel, S.~B., Guyon, O., Hall, D. N.~B., Jovanovic, N., and Atkinson, D.~E.,
  ``Evolutionary timescales of {AO}-produced speckles at {NIR} wavelengths,''
  in [{\em Adaptive Optics Systems V 9909E}{\nolinebreak\hspace{0.1em}]},  {\em
  Proc. SPIE} {\bf 9909},  9909--46 (2016).

\bibitem{Vanderburg16a}
Vanderburg, A., Becker, J.~C., Kristiansen, M.~H., et~al., ``Five planets
  transiting a ninth magnitude star,'' {\em ApJ Letters} (in press),
  arXiv:1606.08441 (2016).

\bibitem{Vanderburg16b}
Vanderburg, A., Bieryla, A., Duev, D.~A., et~al., ``Two small planets
  transiting {HD} 3167,'' {\em ApJ Letters} (in press),  arXiv:1607.05248
  (2016).

\end{thebibliography}
\bibliographystyle{spiebib}   %>>>> makes bibtex use spiebib.bst

  \end{document}